# Synthesis of $V_2O_3$ Nanoplates

Hamid Reza Rasouli[1], Naveed Mehmood[1], Onur Çakıroğlu[2], Engin Can Sürmeli[1], T. Serkan Kasırga[1,2]

[1] Bilkent University UNAM – National Nanotechnology Research Center, Ankara, Turkey 06800

[2] Department of Physics, Bilkent University, Ankara, Turkey 06800

**Abstract:** Peculiar features exist in the stress-temperature phase stability diagram of $V_2O_3$ such as a first order phase transition between the paramagnetic insulating and metallic phases that ends with a critical point, quantum phase transition, and a triple point. These features remain largely unexplored and the exact nature of the phase transitions is not clear due to very limited control over the stress in bulk or film samples. Here, we show the synthesis of single-crystal $V_2O_3$ nanoplates for the first time using chemical vapor deposition via van der Waals epitaxy. Thickness of the $V_2O_3$ nanoplates range from a few to hundreds of nanometres and they can be mechanically exfoliated from the growth substrate. Using Raman spectroscopy on the nanoplates, we reveal that upon heating, $V_2O_3$ enters supercritical state for both tensiley strained and relaxed crystals with similar out-of-plane response. Transmission electron microscopy on $V_2O_3$ nanoplates hints the existence of a structural change when the crystals are heated. Our results show that $V_2O_3$ nanoplates should be useful for studying the physics of supercritical state and the phase stability of $V_2O_3$ to enable new horizons in applications.

**Main text:** Vanadium oxides display peculiar properties due to the unpaired d-electrons in vanadium's valance. Interplay among the internal degrees of freedom such as spin, charge and orbital leads to rich electronic phenomena such as metal-insulator transition (MIT)[1], unusual quantum spin states[2,3] and superconductivity[4]. These phenomena are remarkably sensitive to the external stimuli such as temperature, stress and doping in vanadium oxides. Thus studying them in bulk or film samples results in multiple problems such as various phase and domain formations due to non-uniform strain, and compositional variations due to local doping and oxygen vacancies[5,6]. These problems can be avoided by the use of crystals that are smaller than the characteristic domain size[7–9]. Free-standing $VO_2$ nanocrystals, as an instance, have been used to determine the metal-insulator phase stability diagram and led to observation of a solid-state triple point[10].

At the room temperature, $V_2O_3$ is in the paramagnetic metallic corundum (PM) phase, while at about 160 K it turns into the antiferromagnetic insulating monoclinic (AFI) phase[1,11]. Furthermore, tensile stress stabilizes the paramagnetic insulating corundum (PI) phase at the room temperature or above[12]. Besides the first order phase transitions, when the crystal is under tensile stress at elevated temperatures (~400 K), PI and PM phase equilibrium ends with the solid-state critical point to enter a solid-state supercritical state analogous to supercritical fluids. A depiction of the stress-temperature phase stability diagram of $V_2O_3$ is given in **Figure 1a** [11]. Attempts to explore the MITs and the criticality in $V_2O_3$ have been mainly focused on chromium doped samples as PI to PM transition can be obtained without tensiley straining the crystal[12–15]. When $V_2O_3$ is doped by a certain amount of Cr, PI phase becomes stable in unstrained crystals. However, this simple minded approach leads to poor metallicity[15] due to coexistence of the metallic and insulating domains stabilized by structural defects[16,17].

Furthermore, the MIT is shown to be inequivalent to the one observed via application of compressive stress[18–20]. Further complications arise in film samples due to polycrystallinity of the films and nonuniform stress due to the substrate adhesion as well as lattice mismatch induced stresses in epitaxially grown films. Similarly, bulk crystals of $V_2O_3$ present other challenges such as irreproducibility in measured quantities[21,22]. So far, no free-standing high-quality crystals smaller than the characteristic domain size were available for $V_2O_3$ due to the difficulties in the synthesis. Thus, the tremendous potential to study the criticality of the Mott transition has not been fully realized[14] and the precise details of the phase diagram is still unclear[23]. In this paper, we show a route to synthesize high-quality layered $V_2O_3$ single crystals of a few nanometer thickness and study some novel aspects of the phase transitions.

Our crystal synthesis method relies on salt-assisted chemical vapor deposition of nanoplates. Potassium Iodide (KI) salt is reacted with $V_2O_5$ at an elevated temperature to form potassium vanadate ($KV_3O_8$) (see supporting information)[24], an intermediate compound that can be further reduced to $V_2O_3$ with $H_2$. This chemical route results in much higher quality crystals as compared to the direct reduction of $V_2O_5$ via $H_2$ and sulphur[25] or reduction of $V_2O_5$ in molten potassium fluoride (KF)[26]. Our synthesis takes place in a custom-made chemical vapor deposition chamber we reported earlier, that allows real time optical observation and control of the synthesis[27]. A mixture of KI:$V_2O_5$ is milled together in the weight ratio of 2:1 and a few fine granules are placed directly on a c-cut sapphire substrate. Then the chamber is vacuumed to $10^{-3}$ mbar and flushed with Ar several times. The growth takes place at an atmospheric pressure of Ar. Once 660 °C is reached, the precursor mixture starts melting and $H_2$ gas flow is turned on and the stage temperature is ramped up to 850 °C. 10 minutes after the introduction of the $H_2$ gas, crystals of varying thicknesses form from the supersaturated droplets in a period of 5 minutes. Once all the droplets are consumed, $H_2$ flow is shut down and the chamber is cooled down at a controlled rate (40 °C/min.). A series of optical microscope images taken during the growth given in **Figure 1b** show the formation of the crystals. Although real time observation of the synthesis allowed us to determine the growth conditions leading to $V_2O_3$ nanoplates and prepare the growth recipes, any conventional split tube furnace can be used for the nanoplate synthesis.

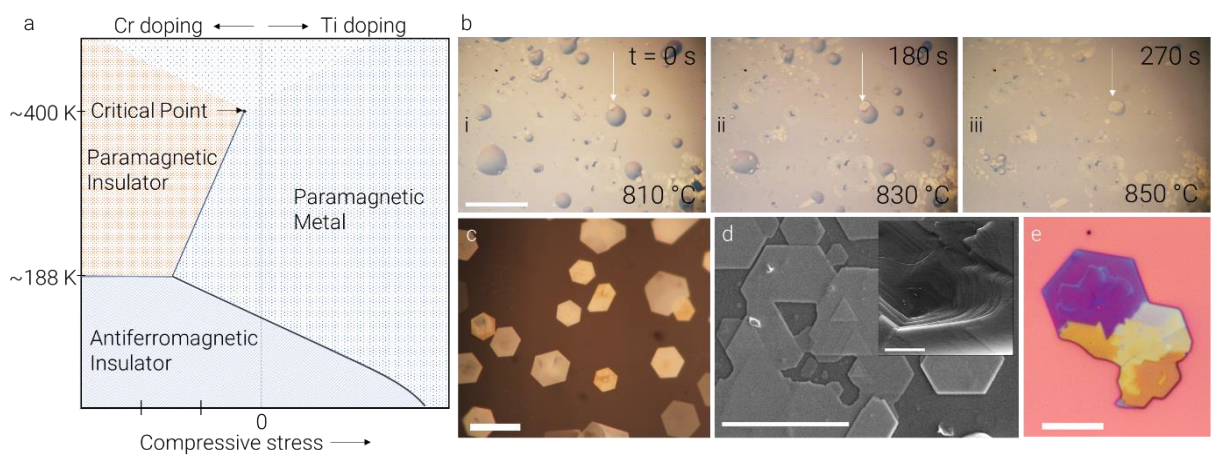

**Figure 1 a** Tentative phase stability diagram of $V_2O_3$ is depicted in the figure based on the previous reports. The effect of doping and stress is illustrated on the lateral axis and the temperature on the vertical axis. Marked temperatures on the graph designate the temperature of the critical point and the triple point. Shaded region above 400 K, beyond the critical point, depicts the supercritical state, where Mott criticality is prevalent. Markings on the horizontal axis indicate 4 kbar/division hydrostatic pressure. **b** Optical microscope micrographs taken during the salt assisted synthesis shows the crystal formation stages. Images are taken from 810 to 850 °C and the beginning of the observation in **i** is marked as time zero (t=0 s). As the time passes by (**ii-iii**), thin crystals nucleate from the liquid flux. White arrow indicates an exemplary crystal formation. Scale bar is 50 µm and subsequent images share the same scale bar. **c** Optical microscope micrograph shows crystal formations of various thicknesses on sapphire. Scale bar is 20 µm. **d** SEM micrograph shows layer-by-layer growth and the inset is a close-up view of the edge of a crystal. Scale bars are 20 µm for the main figure and 1 µm for the inset. **e** Optical microscope micrograph of a $V_2O_3$ nanoplate transferred on to oxidized Si chip showing the crystal mechanically exfoliated from the substrate. Scale bar is 10 µm.

Optical microscope micrographs show crystals of varying thicknesses (**Figure 1c**). Images obtained by scanning electron microscopy (SEM) given in **Figure 1d** and in its inset reveals layer-by-layer growth of the crystals. This mechanism is commonly observed in the crystal formation from supersaturated liquid precursors[28]. As the solute particles energetically favour to attach to kinks rather than flat surfaces, such steps occur. Strikingly, these layers can be exfoliated mechanically. When we transfer crystals from the growth substrate onto $SiO_2$/Si substrate using cellulose acetate butyrate (CAB), transferred crystals typically have a few nanometre thin layers attached to the thicker crystals. An example is shown in **Figure 1e**. Such exfoliation from the bulk is observed in both van der Waals stacked 2D layered materials and non-van der Waals stacked materials: Hematite (α-$Fe_2O_3$), which shares the same crystal structure with the metallic phase of $V_2O_3$, and ilmenite ($FeTiO_3$) 2D layers have been obtained via liquid phase exfoliation[29,30]. Atomic force microscope (AFM) scans show that the minimum step height among the layers is 0.6 nm (see supporting information).

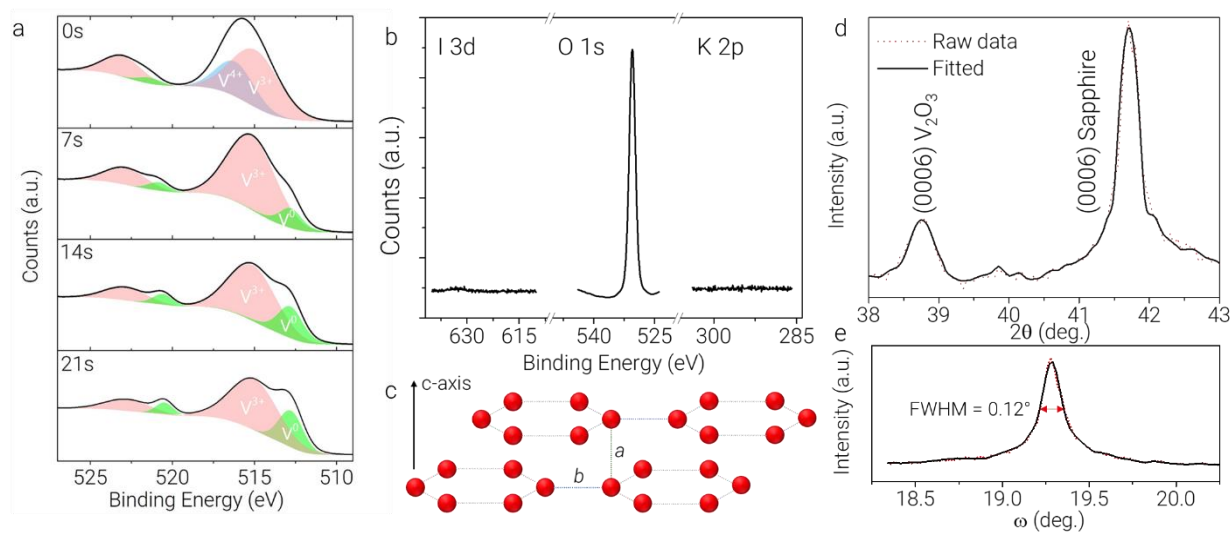

**Figure 2 a** XPS spectra of the V 2*p* peaks taken from pristine samples and samples etched for several seconds. Fits to the spectra by V 3+, 4+ and metallic states are given by pink, blue and green filled curves, respectively. **b** XPS surveys of I 3*d*, O 1*s* and, K 2*p* peaks indicate no I and K presence. O 1*s* peak is dominated by the binding of O in the substrate as the surface coverage of $V_2O_3$ crystals are low. **c** Sketch of the arrangement of vanadium atoms in the corundum phase. *a* and *b* indicate the lattice parameters along and perpendicular to the c-axis, respectively. Upon transition from PM to AFI phase, *a* expands by ~1.9% while *b* expands by ~4.2%. A similar expansion takes place in transition from PM to PI phase. **d** XRD scan shows the (0006) planes of $V_2O_3$ and c-cut sapphire (ICDD card no. 04-002-9108). **e** Rocking curve scan gives a single peak for $V_2O_3$ (0006) showing that the c-axis of the corundum $V_2O_3$ is parallel to the substrate surface.

To confirm the chemical composition of the synthesized crystals, we performed wavelength dispersive X-ray spectroscopy (WDS) and X-ray photoelectron spectroscopy (XPS). WDS spectra taken from a few tens of nm thick crystals show only V and O in the crystals (see supporting information). Consistent with the WDS, XPS surveys show that there are O, V and Al yet no I and K for the samples when rinsed with deionized water (**Figure 2a-b**). For unrinsed samples, K peak is observed in the XPS spectrum, however, after 1 second etch with 1 keV argon ions, K peak disappears, indicating remnants of the reaction by-product on the sample surface. We don't observe potassium in the crystals as determined by further depth profiling with XPS via Ar ion milling. This is not surprising as it is expected that K atoms would be excluded from the $V_2O_3$ lattice due to their large size[12]. Al $2p$ and O $1s$ peaks are due to the sapphire substrate. XPS spectra for V $2p$ peaks confirm that vanadium is in 3+ valance state with a slight native oxide formation that can be removed by Ar ion milling as shown in **Figure 2a** [31]. Excessive etching results in the formation of metallic vanadium on the surface. However, exposure to air at elevated temperatures or for prolonged periods oxidizes the crystals even further and the secondary peak that belongs to 4+ valance state in V $2p$ spectrum becomes more pronounced.

X-ray diffraction (XRD) confirms the c-axis of the sapphire substrate and the synthesized crystals are parallel to each other. **Figure 2c** depicts the vanadium atoms in the corundum phase. Upon transition from PM to AFI phase, V-V distance along the c-axis (*a*) increases by ~1.9% and ~4.2% in-plane (*b*). Similarly, transition from PM to PI phase increases the V-V distance by ~1.9% along the c-axis and ~1% in-plane[32,33]. θ-2θ scans around the (0006) reflections of the sapphire and $V_2O_3$ shows that the corundum c-axis is orientated perpendicular to surface (**Figure 2d**). We assign the peaks at 41.7° to (0006) plane of the sapphire substrate as they are missing in the grazing incidence XRD diffraction. The remaining peak at 38.7° matches very well with those reported for $V_2O_3$ on α-$Al_2O_3$ (ICDD card no. 04-002-9108). Rocking curve scan (ω-scan) of the $V_2O_3$ (0006) reflection results in a very sharp peak with a full width half maximum (FWHM) = 0.12°. This shows corundum c-axis of the crystals is perpendicular to the c-plane sapphire surface. We consider that nanoplate growth on the sapphire surface to be an incommensurate van der Waals epitaxy[34]. Lattice mismatch between the c-cut sapphire and $V_2O_3$ would result in 4.1% biaxial in-plane tensile strain. However, as we discuss in the following paragraph this level of strain is unlikely[35]. We would like to emphasize one more time that $V_2O_3$ crystals can be mechanically removed from the substrate surface. Another observation that supports the incommensurate van der Waals epitaxy is our failed attempts to grow $V_2O_3$ crystals on various substrates such as oxidized Si, Si and quartz.

Raman spectrum of PM $V_2O_3$ has six well defined peaks corresponding to $E_g$ and $A_{1g}$ vibrational modes at 203, 290, 590 $cm^{-1}$ and 231, 500 $cm^{-1}$, respectively[36,37]. Raman spectra for the as-grown crystals on sapphire have peaks around 248, 305 and, 514 $cm^{-1}$ and they match well with the reported spectrum of the paramagnetic insulating phase[37]. Due to the thermal expansion coefficient (TEC) difference between the sapphire and $V_2O_3$, when the crystals cooled down to the room temperature, they are under tensile stress[38]. Indeed, from the temperature where crystals form (~850 °C) down to the room temperature, TEC difference perpendicular to the corundum c-axis accounts for ~1.1% tensile strain on the crystals (details of the calculations are provided in the supporting information). This corresponds to 0.29% volumetric expansion or -6.6 kbar hydrostatic pressure[12]. In McWhan's phase stability diagram[11], such negative pressure is enough to stabilize PI phase at room temperature. The mismatch between TECs also implies that the growth temperature has an influence on the strain at room temperature. Lower growth temperature will result in larger effective hydrostatic pressure. As we discuss in the following paragraphs, these observations are consistent with the fact that as synthesized crystals are in the paramagnetic insulating phase.

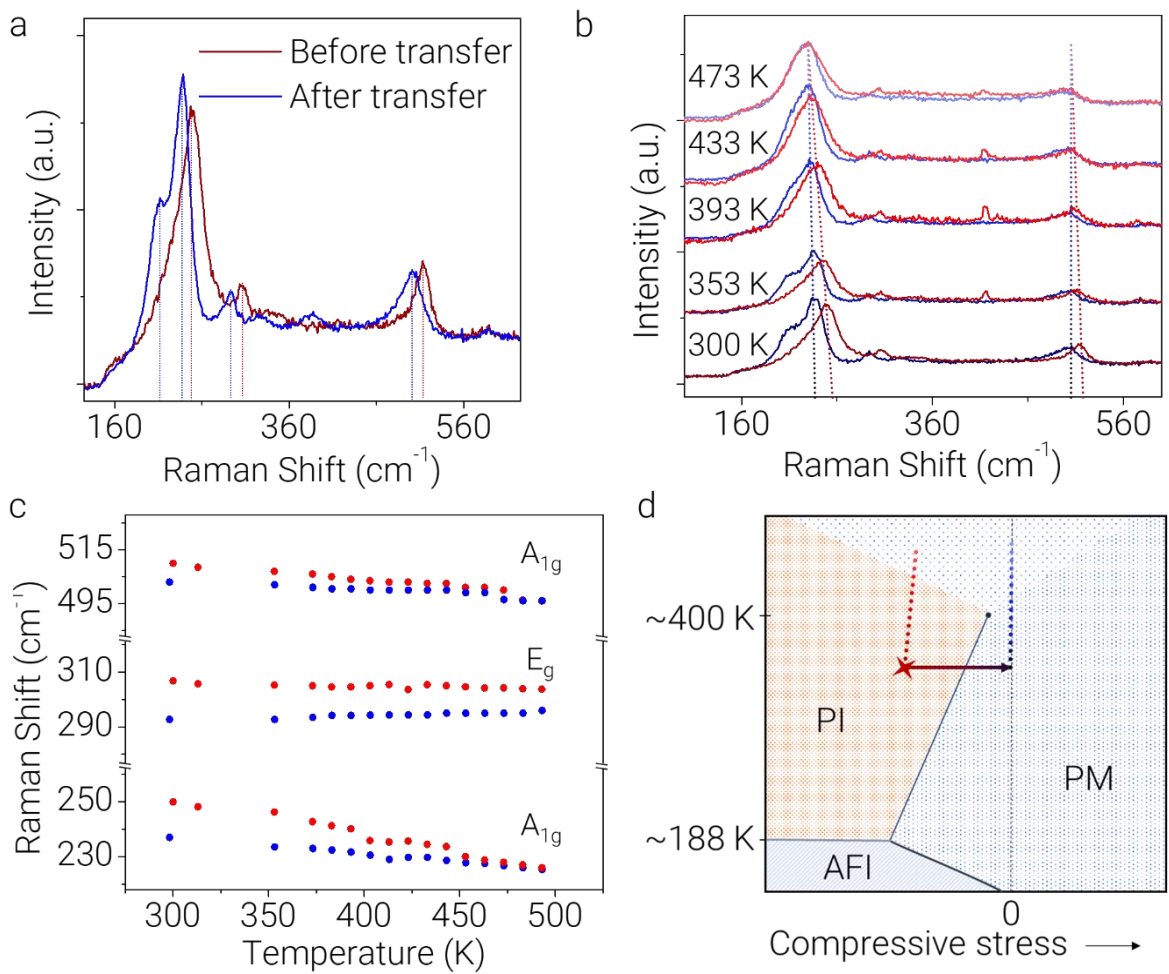

**Figure 3 a** Raman spectra of as-grown and transferred $V_2O_3$ on sapphire are given in the figure. Major Raman modes are marked by dashed lines with respective colours. **b** Raman spectra of as-grown (red series) and transferred (blue series) crystals become similar upon heating. $E_g$ mode around 210 cm$^{-1}$ gradually disappears with the increasing temperature and $A_{1g}$ mode softens in both cases to merge around 450 K. The peak around 416 cm$^{-1}$ is due to the sapphire substrate. **c** Temperature dependence of the peak Raman modes for as-grown (red points) and transferred (blue points) crystals are given in the figure. Raman peak positions can be determined with ±0.5 cm$^{-1}$ precision. **d** Effects of releasing the crystals and heating both as-grown and transferred crystals are illustrated on the tentative phase stability diagram. Red star marks the tentative stress on the as-grown nanoplate at room temperature and the solid arrow emerging from it indicates the effect of transferring. Red and blue dotted lines depict how both as-grown and transferred crystals enter supercritical state, respectively. Red dotted line is drawn slanted as the temperature increase will reduce the tensile stress due to differences in thermal expansions of $V_2O_3$ and the substrate.

A striking change takes place in the Raman spectrum after releasing crystals from the sapphire surface or when the crystals are agitated strongly on the sapphire surface. **Figure 3a** shows the Raman spectrum of a crystal transferred (see experimental methods for details) onto a clean sapphire substrate and a crystal on the growth substrate for comparison. Peak positions of the transferred crystals match very well with those of the PM phase reported earlier[36,37].

This change in the spectrum can be explained by the phase transition from PI to PM phase due to the relaxation of the tensile stress on the crystals.

When we heat $V_2O_3$ thin crystals, we observe reversible softening of the Raman modes especially above 400 K (**Figure 3b**). Intriguingly, crystals both as-grown and transferred to a pristine substrate show the similar Raman features above 400 K. Cooling back to the room temperature retrieves the initial spectra for both the phases. This observation can be explained by emergence of the supercritical state upon heating $V_2O_3$. As PI and PM phase boundary ends in a second-order critical end-point[14], crystals heated into the supercritical state display a unique Raman spectra irrespective of their initial phase. **Figure 3c** shows the temperature dependence of Raman peaks that persists up to the supercritical state for both PI and PM crystals. $A_{1g}$ peaks for both PI and PM phases merge at elevated temperatures while the $E_g$ peaks show little variation with the temperature. This observation can be explained by insensitivity of in-plane $E_g$ modes to the second order phase transition since $e_g(\pi)$ band is perpendicular to the c-axis[39] and motion of atoms along the c-axis mainly effects the $A_{1g}$ modes[37]. **Figure 3d** depicts how different initial phases end up in similar Raman responses on a tentative phase stability diagram. This point is also supported by the asymmetry observed in the 514 cm$^{-1}$ peak at the supercritical state. The asymmetric feature can be attributed to the Fano resonances of Raman active electronic transitions across $a_{1g}$ and $e_g(\pi)$ bands near the Fermi level[40]. As the number of electrons in the $a_{1g}$ band increases in the supercritical state compared to the PI phase, continuum electronic Raman process belonging to the same symmetry group as the phonon Raman scattering due to $A_{1g}$ phonons may interact with each other. Interaction between a discrete process with a continuum process results in an asymmetry in the Raman peak.

To further characterize the crystal structure and crystallinity of the nanoplates we performed transmission electron microscopy (TEM) and selected area electron diffraction (SAED) on $V_2O_3$ nanoplates transferred to holey carbon and silicon nitrite TEM grids (**Figure 4a-d**). We also took high-resolution TEM images and SAED patterns from the cross-section of the $V_2O_3$ on sapphire (**Figure 4e-f**). When the zone axis is perpendicular to the sample surface, we find a six-fold symmetric arrangement with d-spacing of 2.5 Å for $(10\bar{1}0)$ plane. This is consistent with the XRD measurements as the corundum c-axis points perpendicular to the nanoplate surface. However, SAED pattern taken along different zone axes match exactly with a face centered cubic (FCC) structure rather than more commonly reported hexagonal closed pack (HCP) corundum structure. **Figure 4b-d** shows SAED patterns taken through various zone axes and angles between various planes are measured. This measurement is confirmed in five separate crystals transferred via different methods (see experimental methods).

Most plausible explanation for the observed shift in stacking from HCP to FCC is a slight structural change upon heating of the crystals by the electron beam to the supercritical state. A temperature rise of ~150 K is sufficient for emergence of the supercritical state in the nanoplates. Indeed, as the crystals are suspended on the TEM grid with minimal thermal contact with the environment, we measure a temperature increase of up to ~4 K per µW of laser power in Raman spectroscopy (see supporting information for the Raman spectra taken with various laser powers on samples prepared for TEM). In a recent study, electron beam heating in TEM is measured using the MIT in $VO_2$ nanocrystals, and even with very modest electron flux it is possible to induce significant temperature rise in the cantilevered nanocrystals[41]. Another evidence that supports the electron beam induced heating to cause the structural change is from the SAED pattern taken from the cross-sectional TEM samples (**Figure 4e-f**). As the $V_2O_3$ crystal is sandwiched between the sapphire substrate and the platinum top layer, it is not thermally insulated from the support structures. Thus, we claim that the beam induced heating in this case is not high enough to cause the HCP to FCC transition. SAED pattern taken from the cross-sectional sample matches very well with the HCP corundum structure (**Figure 4f**). We would like to note that there are no previous reports of FCC structure for the high-temperature $V_2O_3$ and further studies are required to elaborate this intriguing observation. In parallel with our Raman and SAED measurements, an anomalous increase in the resistivity of the PM phase[21,42] and a minima in the c-axis lattice parameter[38] around 400 K is observed. This indicates emergence of a new charge ordering.

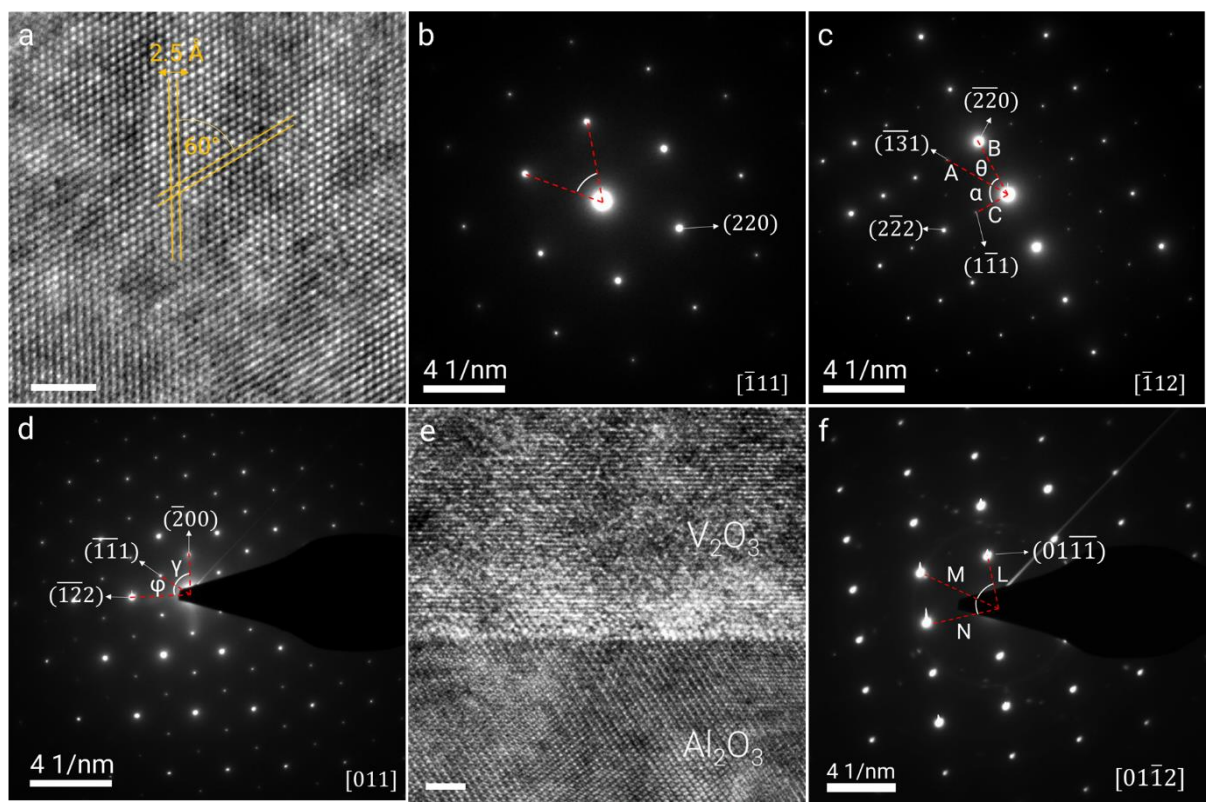

**Figure 4 a** HR-TEM image of a $V_2O_3$ crystal. Close-packed structure is evident. Scale bar is 2 nm. **b** Six-fold symmetric SAED pattern indexed according to FCC [$\bar{1}$11] direction. However, same pattern is produced for HCP [0001] direction. Length of the red dashed lines are 4 nm$^{-1}$, corresponding to 2.5 Å d-spacing along (220) FCC or (10$\bar{1}$0) HCP planes. **c** Once the sample is tilted to a different zone axis, the difference between the FCC and HCP structures becomes clear. No such pattern can be found for HCP. Angles between the principle plane normals $\theta = 39.23°$, $\alpha = 50.77°$ and the ratios of the principal spot spacings B/C = 1.633, A/C = 1.915 match exactly with the standard indexed diffraction pattern for FCC crystals. **d** Another specific FCC diffraction pattern is obtained and indexed along [011] direction. Angles between the principle plane normals $\gamma = 54.74°$ and $\varphi = 35.26°$ matches perfectly with the standard indexed diffraction pattern for FCC crystals. **e** Cross-sectional HR-TEM image shows the stacking of $V_2O_3$ crystal on $Al_2O_3$ substrate. Scale bar is 2 nm. **f** SAED pattern taken from $V_2O_3$ sample prepared for the cross-sectional TEM imaging can be uniquely indexed with HCP [01$\bar{1}$2] diffraction pattern. Both the angles between the principle plane normals and N/L = 1.52, M/L = 1.82 matches very well with the standard indexed SAED for HCP.

Now, we turn our attention to the electrical properties of the nanoplates. First, we patterned as-grown crystals with Au/Cr contacts. Details of the device fabrication are given in the experimental methods section. Resistance vs. temperature (RT) curve taken using four-terminal configuration from 150 K to 300 K shows activated electrical conduction throughout the entire temperature regime with a rapid increase in the resistivity below 188 K (**Figure 5a**). This is consistent with the fact that our as-grown crystals are in PI phase and as they cool down, PI to AFI transition takes place at the previously reported transition temperature[11]. Phase transition is not sharp since as-grown crystals are under non-uniform stress due to the surface adhesion leading to the coexistence of PI and AFI phases over a temperature range. Resistivity ($\rho(T)$) vs. temperature ($T$) curve below 188 K, except the vicinity of the phase transition temperature, can be fitted with an Arrhenius relation $\rho(T) = \rho_0 e^{\frac{E_A}{k_B T}}$ with $E_A = 0.195 \pm 0.005$ eV activation energy, $k_B$ is the Boltzmann's constant and $\rho_0$ is a constant [21].

When we transfer the nanoplates from the growth substrate to an oxidized Si chip, tensile stress on the nanoplate is released and the nanoplate turns to PM phase. We observe the metallic resistivity down to ~160 K (**Figure 5b**). At 160 K, MIT begins. MIT is not abrupt in our devices since the nanoplate still adheres to the $SiO_2$ surface. Moreover, due to the clamping effect of the electrical contacts, some force is exerted on the nanoplate upon the onset of phase transition since the in-plane lattice constant increases significantly during the MIT. This results in a coexistence of the metallic and the insulating phases over a temperature range and produce a staircase-like change in the resistance with the temperature[43]. Based on the RT measurements with known crystal thicknesses, we calculated the temperature dependent resistivity ($\rho(T)$) of the metallic phase as $\rho(T)/\rho(273\ K) = 0.58 + 1.510^{-3}T$. $\rho(T)/\rho(273\ K)$ is comparable to the bulk samples[42,44].

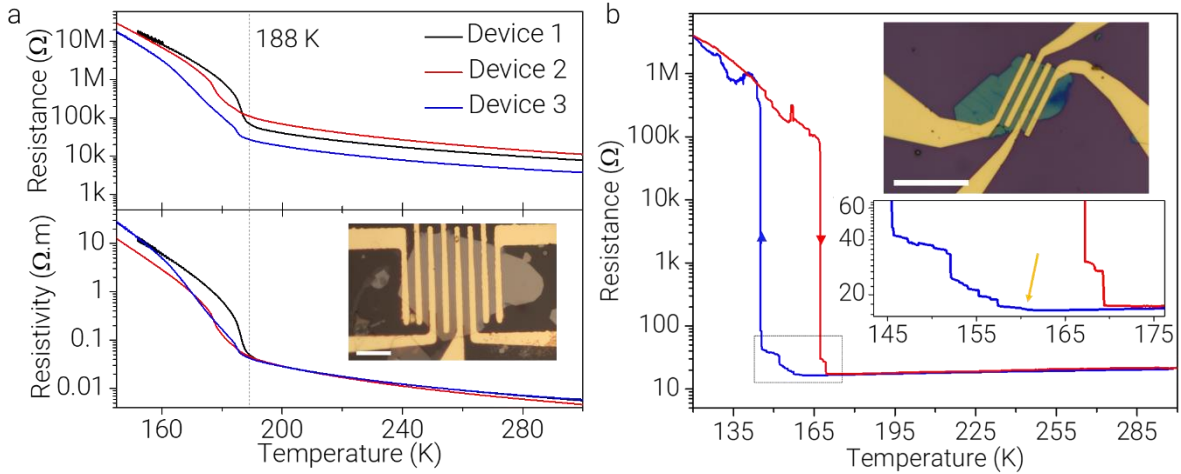

**Figure 5 a** Resistance and resistivity vs. temperature curves for three different devices are given in the graph. All devices show insulator to insulator transition around 188 K. Inset in the lower panel shows a typical as-grown nanoplate patterned using optical lithography. Scale bar is 20 μm. **b** Resistance vs. temperature curve for a transferred $V_2O_3$ crystal. For this particular sample there is a 5 ½ orders of magnitude change in the resistivity. Here, blue and red colours represent cooling and heating, respectively. Lower inset shows a scale up of the region marked with dashed rectangle. Onset of the phase transition is around 160 K. Upper inset shows a typical device patterned using electron beam lithography.

High-quality $V_2O_3$ nanoplates that we introduce in this work can be both tensiley and compressively stressed. This may provide a unique opportunity to study the phase stability diagram of pristine $V_2O_3$. In particular, these crystals may enable a detailed study of supercritical state and cross-over along the Widom line[45] that separates the Mott insulator like and correlated metal like regions. FCC structure we observed under electron beam imaging requires further investigation as the existence of various phases may provide further insight towards theoretical understanding of the supercritical state. Crystals introduced in this study would be a very suitable test platform for the field-effect experiments on various regions of the phase diagram. These nanoplates can be used in the study of van der Waals-like novel heterostructures for reconfigurable electronics and optoelectronics[46–48].

**Experimental Methods**

Details of the $V_2O_3$ nanoplate synthesis is given both in the main text and the supporting information. XPS analysis is performed using K-Alpha X-ray photoemission spectrometer by Thermo Scientific. X-ray diffraction measurements are taken using XRD Panalytical MRD X'pert Pro diffractometer. SAED and HR-TEM measurements are taken using FEI Technai G30. As-grown crystals are transferred on holey carbon or Silicon Nitride grids using both micromanipulator and wet transfer methods. In the wet transfer method, growth substrate is coated by a thin Cellulose Acetate Butyrate (CAB) film and baked at 150 °C for 5 minutes. Then, the film is removed via wedging into water and the film is transferred on to target substrate. Finally, the film is dissolved in Ethyl acetate for 15 minutes leaving the crystals on the target[49]. Raman measurements are taken using 532 nm excitation laser focused to a diffraction limited spot. Laser power is kept low enough to restrict photothermal heating. Two and four terminal devices of $V_2O_3$ are prepared using standard optical or electron beam lithography methods. Once the crystals are patterned, they are dipped in buffered oxide etch for a few seconds to remove oxides from the crystal surface for lower contact resistance. Au/Cr layer of ~100/5 nm thickness is deposited using thermal evaporation. No difference observed for devices prepared with Au/Ti of the same thickness with electron beam evaporation.

**Supporting Information**

Experimental details, details of the crystal synthesis, AFM, SEM and additional optical images, further characterization of the crystals. (PDF)

**Corresponding Author**

*E-mail: kasirga@unam.bilkent.edu.tr

**Author Contributions**

T.S.K. conceived the project. H.R.R. and T.S.K. performed the experiments. N.M. and E.C.S. helped with device fabrication and Raman measurements. N.M. performed the crystal transfers. O.Ç. performed the AFM measurements. The manuscript was written through contributions of all authors. All authors have given approval to the final version of the manuscript.

**Funding Sources**

TUBITAK 1001 – Grant No: 116M226

**Acknowledgements**

This work is supported by the Scientific and Technological Research Council of Turkey (TUBITAK) under grant no: 116M226. Authors thank David H. Cobden for useful discussions.

**Notes**

Authors declare no competing financial interests.

# Supporting Information: Synthesis of $V_2O_3$ Nanoplates

Hamid Reza Rasouli[1], Naveed Mehmood[1], Onur Çakıroğlu[2], Engin Can Sürmeli[1], T. Serkan Kasırga[1,2]

[1] Bilkent University UNAM – National Nanotechnology Research Center, Ankara, Turkey 06800

[2] Department of Physics, Bilkent University, Ankara, Turkey 06800

**Synthesis of $V_2O_3$ nanocrystals**

Synthesis steps are outlined in the main text. Here, we would like to show that when $V_2O_5$ reacts with KI, $KV_3O_8$ forms. To illustrate the point, we heated the chamber to 700 °C and watched reaction to take place. Once a droplet is formed, we shut down the furnace and characterized the intermediate compound using energy dispersive X-ray (EDX), Raman and X-ray photoelectron spectroscopy (XPS). EDX maps, **Figure S1**, show the constituent elements of the droplet. K, V and O dominates the spectra with Al signal outside of the droplet coming from the growth substrate. Raman spectrum from the droplet matches very well with $KV_3O_8$ and +5 oxidation state of V from XPS confirms the Raman spectrum.

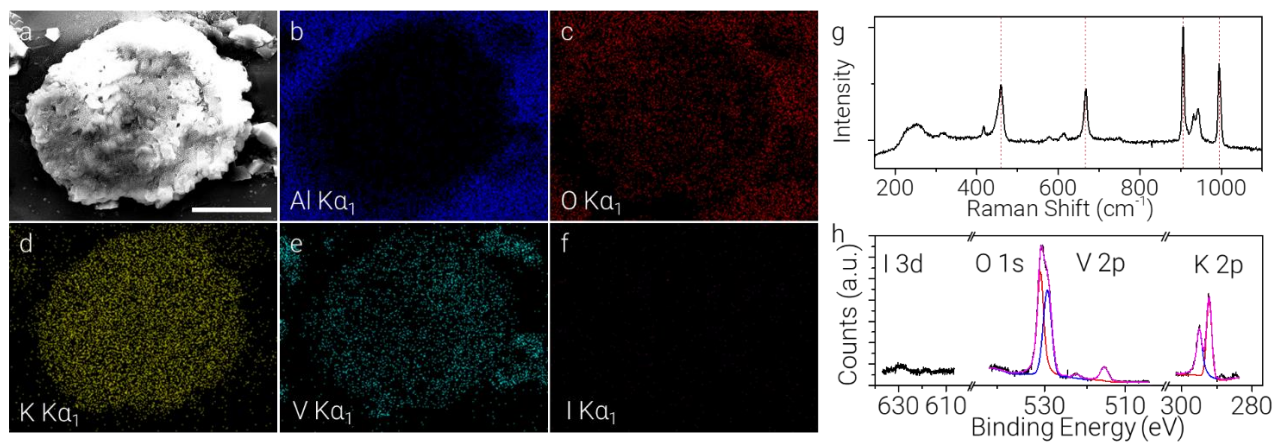

**Figure S 1 a** SEM image of an intermediate compound. Scale bar is 5 µm. **b-f** EDX maps for various elements. The intermediate compound contains O, K and V and no I. Al $K\alpha_1$ signal comes from the sapphire substrate. **g** Raman spectrum taken from the intermediate compound shows typical Raman modes for $KV_3O_8$. These modes are marked by red dashed lines. **h** XPS surveys for I 3d, O 1s, V 2p and K 2p confirms the EDX and Raman sprecta. Fits to the O 1s and V 2p match with the binding energies of $KV_3O_8$.

$V_2O_3$ crystals form via reduction of $KV_3O_8$ under reducing $H_2$ environment. We introduced Se in some trials to form $H_2Se$, a stronger reducing gas, however no significant difference is observed in crystal morphologies. We propose the following reaction for the crystal formation:

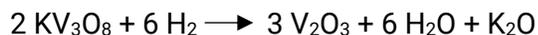

$2\ KV_3O_8 + 6\ H_2 \longrightarrow 3\ V_2O_3 + 6\ H_2O + K_2O$

This proposal is based on the observation that we encounter very small amount of K on the surface of the growth substrate. KOH would evaporate at the growth temperature and $K_2O_2$ readily decomposes into $K_2O$ above 500 °C. Thus, such a reduction route is plausible.

**Atomic force microscopy on $V_2O_3$**

We performed detailed atomic force microscopy measurements on the as-grown and transferred crystals. Some examples are provided in **Figure S2**.

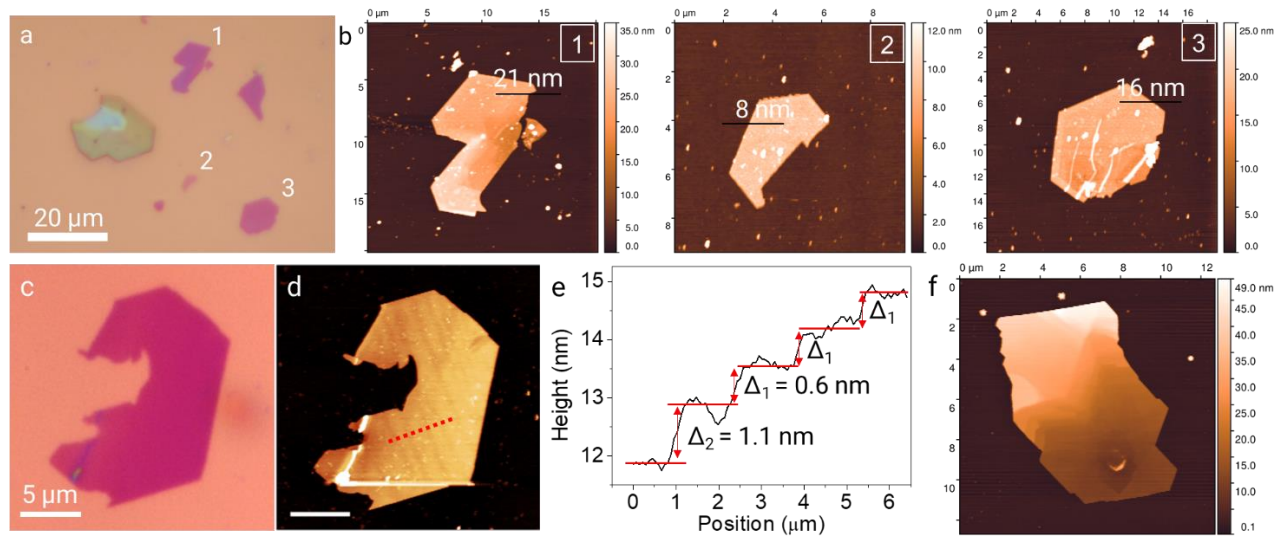

**Figure S2 a** Optical microscope micrograph of transferred thin $V_2O_3$ crystals. AFM maps corresponding to each numbered crystal is given in **b**. **b** Thickness of each crystal is noted on the crystal. Optical contrast among the crystals is not significant. **c** Another crystal transferred on to $SiO_2$ **d** and the corresponding AFM height trace map. **e** Height profile taken along the red dashed line shows steps of 0.6 nm along the sample. **f** AFM height trace map of an as-grown crystal with thickness ranging from 10 to 40 nm. Layered structure is clearly visible from the map.

**Wavelength dispersive spectroscopy (WDS) results**

Wavelength dispersive spectroscopy (WDS) taken from $V_2O_3$ nanoplates shows V, Al and O in abundance. Al and some O results from the substrate. Other marked elements such as K and Se are statistically insignificant in the amount.

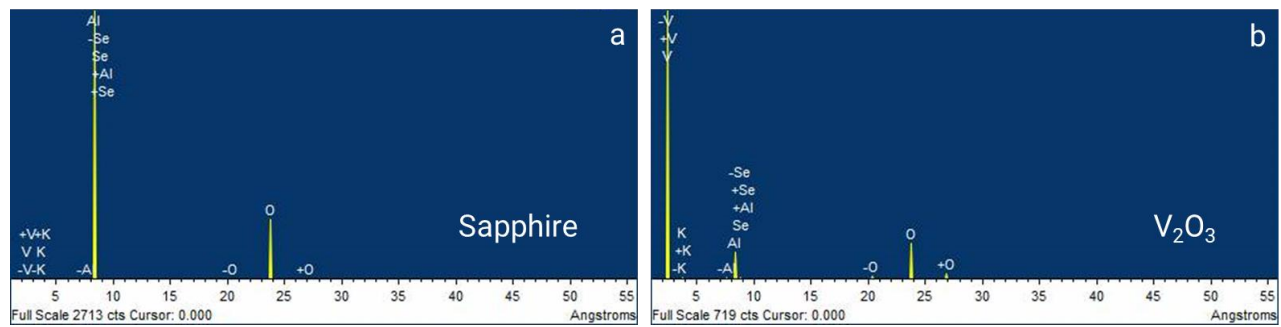

**Figure S3 a** WDS spectrum from the sapphire substrate yields peaks for Al and O. **b** WDS spectrum taken from a $V_2O_3$ crystal shows peaks for V, O and Al. Al and O is due to the substrate.

**Optical images taken at various temperatures**

**Figure S4** shows the room temperature and elevated temperature optical microscope micrographs for both as-grown and transferred crystals. There is no visible optical contrast.

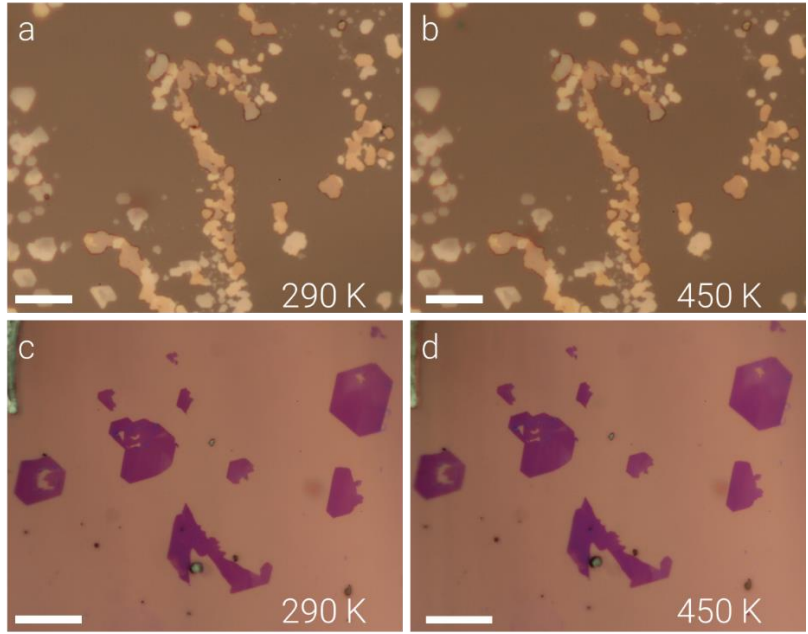

**Figure S4 a** Optical microscope micrograph of as-grown $V_2O_3$ nanoplates at 290 K and **b** at 450 K. Scale bar is 10 µm. **c** Optical microscope micrograph of transferred grown $V_2O_3$ nanoplates at 290 K and **d** at 450 K. Scale bar is 20 µm. There is no optical contrast between the samples in the supercritical state and the respective initial phases.

**Stress on as-grown crystals**

Thermal expansion coefficient (TEC) difference between the sapphire and $V_2O_3$ plays a role in the stress in as-grown crystals. TEC parallel to the c-cut sapphire surface ($\alpha_S^\parallel$) ranges from 8.5 to 5.3 $10^{-6}$ K$^{-1}$ within 850 to 25 °C. $V_2O_3$ has a negative TEC value for 25 to 400 °C and positive above 400 °C along the corundum c-axis ($\alpha_{V_2O_3}^\parallel = -1.273 + 1.782 \ 10^{-8} \ T + 2.405 \ 10^{-13} \ T^2$ °C$^{-1}$) and positive for the a-axis ($\alpha_{V_2O_3}^\perp = 3.266 \ 10^{-5} - 2.871 \ 10^{-8} \ T + 7.747 \ 10^{-13} \ T^2$ °C$^{-1}$) [1]. In plane strain due to thermal expansion coefficient difference will be isotropic. Length change upon cooling from the growth temperature to the room temperature is $\Delta L_S = 5.5 \ 10^{-3} \ L$ for sapphire surface and $\Delta L_{V_2O_3} = 16.7 \ 10^{-3} \ L$ for $V_2O_3$. This difference results in ~1.1% in-plane strain on the crystals. Assuming isotropic Poisson's ratio ($\nu = 0.33$) for[2] $V_2O_3$, total hydrostatic volumetric expansion, $\frac{\Delta V}{V} = \left(1 + \frac{\Delta L}{L}\right)^{1-2\nu} - 1$, would be 0.29%. Isothermal bulk modulus is given by $K_T = -V\frac{dP}{dV} = 1.75$ Mbar for $V_2O_3$ following Sato et. al.[3], and using this relation we can calculate the equivalent hydrostatic pressure as -6.6 kbar. This negative hydrostatic pressure is large enough to stabilize PI phase at room temperature[4]. It is possible to synthesize crystals at lower temperatures, down to 700 °C, as well. Thus, the range of hydrostatic pressure experienced by the crystals due to TEC differences would be from -6.6 to -5.9 kbar.

## Removal of surface oxide for device fabrication

XPS measurements show that samples exposed to air for longer than a day oxidize on the surface. This oxide formation results in poor electrical contact to the crystals, thus needs to be removed prior to the metal contact deposition on to the patterned crystals. We used buffered oxide etchant (BOE) (6:1 volume ratio of 40% $NH_4F$:49% HF in water) to remove the oxide formation from the surface for better electrical contacts. **Figure S5** shows AFM height profiles of a crystal before and after BOE treatment. After 5 seconds of BOE treatment crystal is thinned down by ~7 nm. Further BOE treatment doesn't reduce the crystal thickness but roughness observed after 5 sec. of treatment becomes more severe.

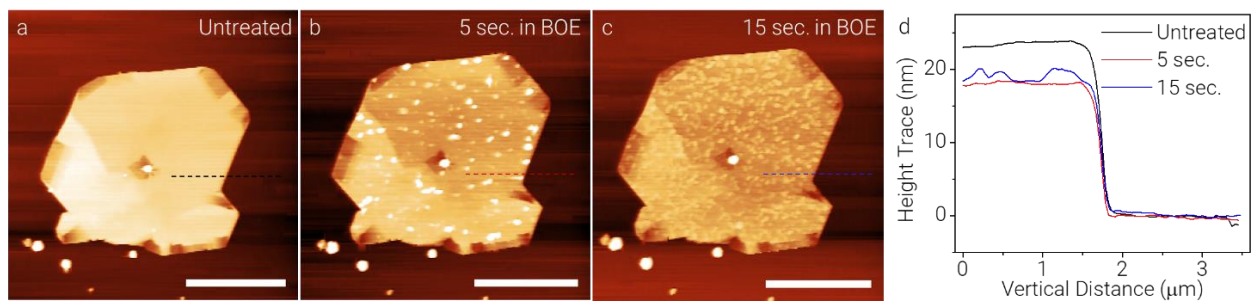

**Figure S 5 a** AFM height trace map of a crystal after several days of exposure to the ambient and **b** same crystal after a 5 second BOE treatment. Surface of the crystal becomes noticeably rougher after the BOE treatment. **c** After 15 seconds of BOE treatment no further reduction in the crystal thickness is observed. However crystal surface is very rough. Scale bars are 4 μm. **d** Height traces taken through the colored dashed lines in respective AFM maps show the change in crystal thickness.

## Mechanical exfoliation of $V_2O_3$ crystals

As grown crystals have limited use for studying various phases of $V_2O_3$. However, these crystals can be mechanically exfoliated from the growth surface. We tried polymer transfer and mechanical removal of crystals using a micromanipulator with success. $V_2O_3$ crystals can be picked up from the growth substrate by a film of 6% polymethyl methacrylate (PMMA) in anisole or 60 mg cellulose acetate butyrate (CAB) in 100 ml ethyl acetate solution[5].

## TEM Samples

TEM samples are prepared both on silicon nitride membrane and holey carbon grids. Transfer techniques described above is used to prepare the samples. Cross-sectional TEM images are taken through samples prepared by focused ion beam (FIB) etching and deposition.

Before TEM studies, we characterize the transferred crystals using Raman spectroscopy to ensure that we observe same Raman profile as the other transferred crystals. First, we notice that crystals suspended on holey carbon grids are exceptionally sensitive to the laser power. This is not too surprising as there is no supporting substrate, absorbed laser power results in a greater local temperature increase in the crystal. Raman spectra taken with ~20, 40 and 80 μW laser powers are given in **Figure S6**. At 20 μW laser power, PM phase is evident. When we apply 40 μW, $A_{1g}$ peak around 240 $cm^{-1}$ shifts towards 230 $cm^{-1}$. Comparing the peak position to **Figure 3c** in the main text we conclude that the temperature rise in the nanoplate is around 4 K/ μW. Energy dispersive X-ray (EDX) spectroscopy taken in TEM shows presence of V and O. Cu and K are also detected in trace amounts. Cu signal results due to scattering of electrons from the Cu grid and K is from the remaining potassium oxide on the crystal surface.

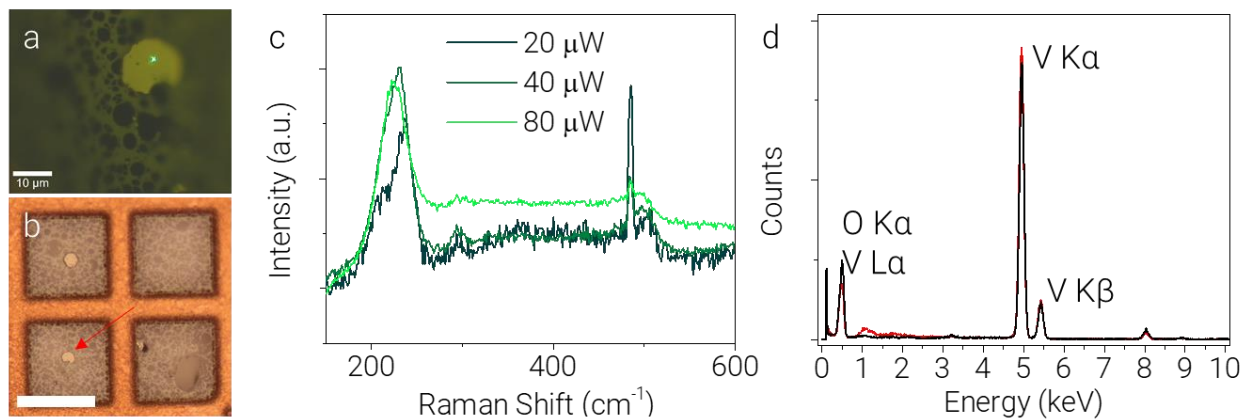

**Figure S 6 a** Optical microscope micrograph of a $V_2O_3$ crystal transferred on holey carbon TEM grid. **b** Smaller magnification optical image of the same crystal in **a** (marked with a red arrow). Scale bar is 50 μm. **c** Raman spectra taken at various laser powers. 20 and 40 μW spectra are scaled accordingly to show the features. Laser heating is apparent as the metallic $V_2O_3$ enters to supercritical state with the increasing laser power. **d** EDX spectrum taken in TEM for two different samples show O and V are the dominant elements. Traces of Cu and K is also observed. Cu is due to the TEM grid and K is the residual oxide on the surface.

### Electrical Measurements

We perform electrical measurements in four-terminal geometry. We source current and measure voltage from the outer terminals and measure the voltage drop from the inner contacts to calculate the resistance across the inner terminals.

Here, we would like to share the resistance-temperature (RT) curve of another device given in **Figure S7a**. In this device, phase transition shows almost 7 orders of magnitude jump in the resistance. All the features observed in the RT curve can be explained qualitatively based on the tentative phase diagram given in the main text (**Figure S7b**). First, we assume that the crystal is relaxed. Although, this may not be true, it doesn't affect the qualitative analysis drawn here. Any residual stress along the crystal would affect the onset temperature of the MIT. When we cool down the sample, like the device reported in the main text, we observe the onset of the transition around 160 K. Then, as the temperature drops further, we observe a staircase like transition indicating coexistence of both phases[6]. This is due to two reasons: First, electrical contacts clamp the nanoplate down and restrict the expansion of the crystal upon MIT and second, the nanoplate adheres to the $SiO_2$ surface resulting a slip-stick expansion of the crystal. Once the whole section in between the electrodes turn to AFI phase, we see the reported activated behavior in resistivity. At this point, the pressure applied by the crystal is great enough to push the contacts to reduce the strain, thus crystal return to the strain-free state. As we start heating the crystal, it superheats up to 170 K. We are not sure exactly why there is a slight increase in the resistance at this point. Then, activated conduction is no longer present and coexistence continues up to 192 K. This temperature is greater than the presumed temperature for the triple point. Thus, some tensile strain on the crystal might be relaxed by the introduction of the PI phase. We would like to remind that PI phase has in-plane lattice constant in between AFI and PM phases. Finally, the resistance returns to its original value abruptly at around 192 K. Each device displays a unique RT curve that can be explained by referring to the phase stability diagram.

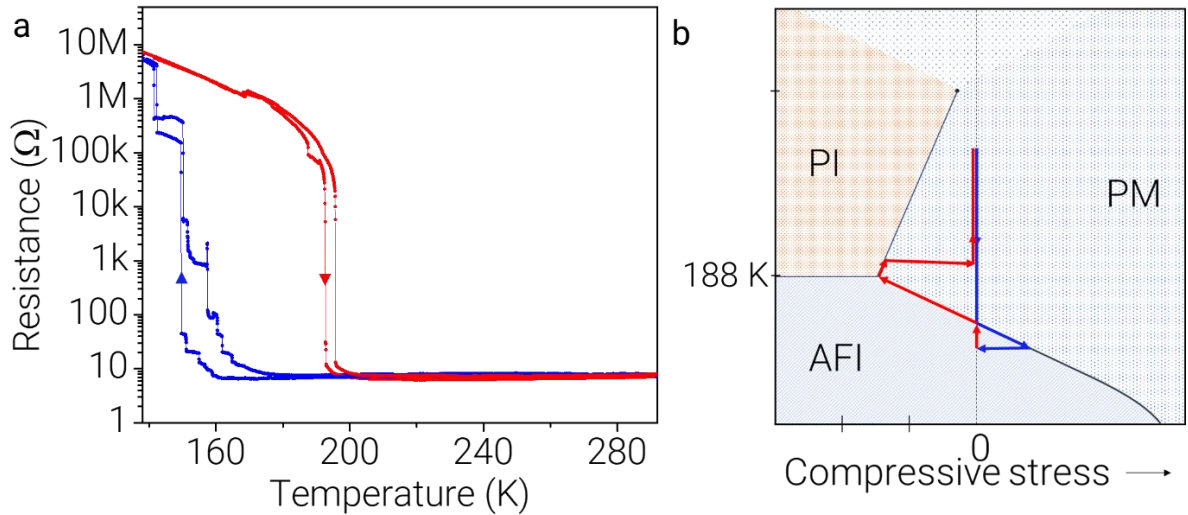

**Figure S7 a** Two RT cycles for a device patterned on a transferred $V_2O_3$ nanoplate. Blue and red colors indicate cooling and heating, respectively. **b** Tentative phase diagram with the colored arrows depicting the changes crystal experiences upon thermal cycling.